# Electronic structure and magnetism in BeO nanotubes induced by boron, carbon and nitrogen impurities.

M.A. Gorbunova,[1] I.R. Shein,[2] Yu.N. Makurin,[1] V.V. Ivanovskaya,[2,3]
V.S. Kijko,[1] A.L. Ivanovskii [2] *

[1] *Ural State Technical University, Ekaterinburg, 620001 Russia*
[2] *Institute of Solid State Chemistry, Ural Branch of the Russian Academy of Sciences, Ekaterinburg, 620041 Russia*
[3] *Laboratoire de Physique des Solides, Uniersity Paris-Sud, CNRS-UMR 8502, 91405, Orsay, France*

(Dated: May, 13, 2008)

We have performed *ab initio* calculations to systematically investigate electronic properties and magnetism of insulating *non-magnetic* beryllium monoxide nanotubes induced by *non-magnetic sp* impurities: boron, carbon and nitrogen. We found that in the presence of these *sp* impurities, which replace oxygen atoms, the non-magnetic BeO NTs transform into magnetic semiconductors, which acquire magnetization caused by spin splitting of (B,C,N) *2p* states located in the forbidden gap of a BeO tube. The magnetic moments of the impurities vary from 0.65 to 1.60 $\mu_B$. On the contrary, when (B,C,N) dopants substitute for Be atoms or in the presence of an oxygen vacancy, the non-magnetic state of the BeO tubes is retained.

## 1. Introduction

At present, diluted magnetic semiconductors (DMSs) such as *d* metal doped III-V (BN, AlN, GaN, GaAs *etc*) and II-VI (ZnO, CdTe, ZnS *etc*) hosts are extensively studied owing to their potential applications in spintronics [1-3]. In this case, the magnetization of the *non-magnetic* III-V or II-VI matrix arises mainly because of the presence of magnetic *d* metal impurities.

An alternative way in the search for potentially useful DMSs is based on a quite unexpected effect of magnetization of *non-magnetic* matrix induced by *non-magnetic sp* impurities, see reviews [4,5]. Moreover, alongside II-VI and III-V crystals, magnetization of some II-VI and III-V nanotubes (such as BN, MgO,

AlN *etc*, see [6-9]) was found which was a result of introduction of *sp* atoms in the tube walls.

Promising II-VI materials include beryllium monoxide BeO – a non-magnetic insulator with a band gap of about 10 eV, which is often used as a refractory material in metallurgy, a material for heat-removing insulators, and a reflector for neutrons and neutron filters; it is also used in laser engineering, ionizing radiation dosimetry, microwave radio-engineering devices, and in other areas [10].

Quite recently, magnetization of wurtzite-like (type *B*4) beryllium monoxide induced by boron, carbon and nitrogen doping was established with the use of *ab initio* band structure calculations [11]. Depending on the type of the dopant, BeO:(B,C,N) systems can behave as magnetic semiconductors or half-metallic magnets. The observed effect is caused by spontaneous spin polarization of the 2*p* bands of boron, carbon or nitrogen impurities which lie above the O2*p* valence band of the matrix. It is quite real to expect the related effects of ***sp-impurity-induced magnetism*** for BeO nanotubular materials.

Very recently, models of pristine beryllium monoxide nanotubes (BeO NTs) were proposed and their structural, cohesion and electronic properties were predicted [12,13]. It was found that these nanotubes are wide-band-gap dielectrics with very small helicity-induced differences in the gaps [12]; in addition, the BeO NTs adopt interesting mechanical properties, namely their Young's moduli are comparable with those for carbon nanotubes [12].

In this Communication, based on DFT-LSDA calculations, we focus on the possibility of improvement of the functional properties of BeO NTs as potential advanced materials and predict the ways for magnetization of *non-magnetic* beryllium monoxide nanotubes by introduction of *non-magnetic sp* impurities.

## 2. Models and computational details

For our investigations of pristine and doped beryllium monoxide nanotubes the following atomic models are chosen. The pristine infinite-long BeO nanotubes are constructed in a conventional way (see, for example [6-8,12,13]) by rolling of BeO graphitic sheets into cylinders. In this way, three groups of BeO-NTs can be created depending on the rolling direction: non-chiral *armchair* (n,n)-, *zigzag* (n,0)- and chiral (n,m)-like tubes.

In our work, we examined two non-chiral nanotubes, namely, *zigzag* (10,0) and *armchair* (6,6) BeO NTs with comparable radii, see Fig. 1. For the simulation of B, C or N doped BeO nanotubes, an oxygen or beryllium atom was removed from the tube wall and replaced by the above *sp* atom. The 160 and 144 atomic cells are used for *zigzag* (10,0) and *armchair* (6,6) BeO NTs, respectively, and the periodic boundary conditions were employed along the nanotube axis. Thus, the *sp* atoms are placed from each other at a distance of about 1.6 nm and may be considered as single (isolated) defects.

Our calculations are based on the density-functional theory and have been performed using the SIESTA code [14]. The valence electrons are described by a linear combination of a numerical atomic-orbital basis set and the atomic core – by norm-conserving pseudopotentials. The pseudopotentials generated using the Troullier and Martins scheme [15] are employed to describe the interaction of valence electrons with the atomic core, and their non-local components are expressed in the full separable form of Kleinman and Bylander [16,17]. The local spin-density approximation (LSDA) formalism [18] is adopted for the exchange-correlation potential. The double-zeta plus polarization atomic-orbital basis set is employed in the calculations. The Hamiltonian matrix elements are calculated by the charge-density projection on a real-space grid with an equivalent plane-wave cutoff energy of 300 Ry. The conjugate gradient algorithm [19] is adopted for the relaxation of the nanotubular structures until the maximum force on a single atom is within 0.01 eV/Å. In view of the periodic boundary condition, our magnetic calculations are performed for the ferromagnetic ordering configuration.

As a result, the equilibrium geometries, atomic magnetic moments (MMs) and the spin-resolved densities of states (DOSs) are obtained and analyzed for above-mentioned doped or vacancy-containing BeO NTs.

## 3. Results and discussion

### *3.1. Pristine BeO nanotubes.*

Let us briefly comment on our data on pristine BeO nanotubes in comparison with earlier findings [12,13]. The initial atomic models of *zigzag* (10,0) and *armchair* (6,6) BeO NTs as depicted in Fig. 1 were constructed by the rolling procedure; these structures have the ideal cylindrical shape, where the interatomic distances are equal to those in the bulk graphite-like hexagonal BeO [21]. At the first step, we performed calculations, in which the initial constructed tubular structures were fully relaxed with respect to atomic positions.

The results obtained lead us to the following conclusions. Firstly, the optimized atomic geometries of the constructed BeO NTs reveal that these systems preserve as a whole the initial cylindrical morphology, which is indicative of the possibility of existence of nanotubular BeO. Secondly, the effect of various relaxations of anions and cations from their ideal atomic position is observed. Namely, the Be atoms move inwards, while the oxygen atoms move outwards with respect to their ideal positions; thus, the optimized geometry of BeO-NTs can be described as two coaxial atomic cylinders with an outer oxygen cylinder and an inner beryllium cylinder. The so-called radial buckling parameter $\beta = R_O - R_{Be}$, where $R_O$ and $R_{Be}$ are the radii of the oxygen and beryllium cylinders, is quite small (0.08 Å) and agrees reasonably with the previously calculated values ($\beta$ = 0.10-0.06 Å) [13].

The ground state of pristine BeO NTs is non-magnetic. Our calculations show that the completely occupied valence zone for Be NTs contains two separate bands A and B, Fig. 2. These bands are formed mainly by 2*s* (band A) and 2*p* orbitals of oxygen (band B) with a small admixture of Be 2*s*, 2*p* states suggesting the predominant ionic bonding between oxygen and Be atoms. The valence zone is separated from the free Be2*p*-like conduction band (band C, Fig. 2) by a band gap (BG) at about 5.9 eV. This value is comparable with earlier estimations in DFT-LDA [12] (BG ~ 5 eV) and DFT-SIC [13] (BG ~ 8 eV) approaches. Note also that there are no significant differences in the *armchair* and *zigzag* configurations for the considered BeO NTs.

### *3.2. Electronic properties and magnetism in BeO nanotubes induced by boron, carbon or nitrogen impurities.*

Let us discuss the electronic properties and magnetism in B, C and N doped BeO tubes. Firstly, the geometry optimization does not lead to any dramatic distortion of the initial BeO NTs structure; this means that the doped BeO tubes are stable and their cylinder-like morphology is retained. Secondly, our calculations show that as a result of replacement of beryllium atom by all the above *sp* impurities these systems remain non-magnetic (as was predicted also for the crystalline BeO [7]), and they will not be discussed further.

The picture is quite different when *sp* dopants are placed in the oxygen sites, and the doped BeO:(B,C,N) NTs tubes become magnetic. In this case, the total magnetic moments (MM, per cell) in the doped BeO tubes vary from 2 to 1 $\mu_B$ (see Table 1). At the same time, the maximal (1.6 - 0.75 $\mu_B$) magnetic moments are localized on the impurity centers, whereas the magnetic moments of the neighboring beryllium (0.12 - 0.01 $\mu_B$) and oxygen (0.06 - 0.01 $\mu_B$) atoms are much smaller.

It is interesting to note that in the sequence of doped nanotubes BeO:B → BeO:C → BeO:N, the total MMs and magnetic moments of *sp* dopants vary non-monotonically adopting the maximal values for the BeO:C tube. This situation differs drastically from the bulk BeO [7], when for the crystalline beryllium oxide lowering of MMs occurs in the sequence of dopants B → C → N. Probably, these differences are caused mainly by the features of electronic distributions of four-fold coordinated atoms in wurtzite-like (type *B*4) beryllium monoxide crystal *versus* three-fold coordination of atoms inside BeO tube walls.

The origin of the mentioned effect may be discussed using the DOSs for the doped BeO:(B,C,N) NTs, Fig. 3. The data presented show that the impurity-induced magnetization for BeO nanotubes originates mainly from the spin splitting of states for *sp* impurity atoms. As can be seen from the *l*-projected density of states (Fig. 3), the introduction of all the impurities (B, C and N) in the BeO tube walls leads to:

(i)   occurrence of new impurity 2*p* bands localized in the gap of the BeO NTs;

(ii)  a shift of the Fermi level ($E_F$), which is located in the region of these 2*p*-like impurity bands; and

(iii) splitting of these impurity bands into two spin-resolved bands with various types of filling.

This results in the above magnetization, and non-magnetic BeO tubes become magnetic semiconductors.

The band gaps for doped BeO NTs decrease drastically from 5.9 eV for pristine to 1.1 eV for BeO:B NT, 1.5 eV for BeO:C NT and 1.0 eV for BeO:N NT. Note that the gaps for BeO:B, BeO:C and BeO:N NTs are of quite different types and are formed between the B2$p^{\uparrow}$-B2$p^{\uparrow\downarrow}$ states, C 2$p^{\uparrow}$ - C2$p^{\downarrow}$ states and N2$p^{\downarrow}$ - N2$p^{\downarrow}$ states, respectively, see Fig. 3.

As it was mentioned earlier, in the sequence of doped nanotubes BeO:B → BeO:C → BeO:N, the total MMs and magnetic moments of *sp* dopants vary non-monotonically, whereas the induced magnetization of the nearest neighbor beryllium cations decreases monotonically from 0.12 $\mu_B$ for BeO:B NT to 0.05 $\mu_B$ for BeO:N NT. Moreover, the induced magnetization for Be, O "sublattices" for doped BeO:(B,C,N) NTs varies also monotonically in the sequence of impurities B→ C→ N amounting to about 23%, 19.8% and 10.5% of common magnetization of BeO:B, BeO:C and BeO:N NTs, respectively.

The mentioned peculiarities can be explained if we take into account the mutual positions of the O2*p* valence band of the BeO tube and the 2*p* band of the doping atoms in the energy spectrum of BeO:(B,C,N) NTs. These positions can correlate with the values of the 2*p* orbital energies ($\varepsilon$) of oxygen forming the valence band of BeO NTs ($\varepsilon^{O\ 2p}$ = -17.188 eV) and the impurity atoms: boron ($\varepsilon^{B\ 2p}$ = -8.429 eV), carbon ($\varepsilon^{C\ 2p}$ = -11.788 eV) or nitrogen ($\varepsilon^{N\ 2p}$ = -15.439 eV) [21]. Thus, for the BeO:B NTs with the maximal magnetization of the nearest neighbor beryllium atoms, the B 2$p^{\uparrow\downarrow}$ orbitals are farthest from the matrix O 2*p*-like valence band top, and closest to the Be 2*p*-like conduction band bottom, which promotes their overlapping. On the contrary, for BeO:(C,N) NTs the (C,N) 2$p^{\uparrow\downarrow}$ bands shift to the top edge of the matrix valence band, and these bands are partially overlapped, Fig. 3.

## 4. Conclusions

We have studied the electronic and magnetic properties of B,C,N substitutionally doped *armchair* and *zigzag* configurations of BeO nanotubes using the spin-polarized density functional calculations.

It is found that the electronic and magnetic characteristics of these materials (such as magnetic moments and band gaps) depend strongly on the type of the impurities (and, evidently, on their concentration) and on their positions, and therefore can be controlled. Namely, if beryllium atoms are replaced by the above *sp* impurities, the BeO NT systems remain non-magnetic.

On the contrary, when *sp* dopants are placed in oxygen sites, the effects of the so-called ***sp-impurity-induced magnetism*** are found.

We affirm that these systems, which contain no magnetic transition-metal atoms, undergo a transformation from non-magnetic insulators (pristine BeO nanotubes) to magnetic semiconductors. It is found that the magnetization mechanism for *sp* doped BeO NTs should be attributed to spin splitting of (B,C,N) $2p^{\downarrow\uparrow}$ bands, whereas magnetization of Be and O atoms is relatively small. Moreover, the band gaps for *sp* doped BeO NTs are determined by impurity spin-polarized bands and decrease drastically as compared with pristine BeO tubes - from 5.9 eV to 1.0-1.5 eV.

In conclusion, the present discussion is focused only on single impurity states in the BeO tubes to study the ***impurity-induced magnetism*** whereas numerous issues could be proposed for future studies. For instance, the possible nanotube wall defects (Be or O vacancies) and the effect of ***vacancy-induced magnetism*** (see [22]), as well as the effect of impurity (or vacancy) content variation on the magnetic properties of these nanotubes may be of interest. Such studies will make it possible to compare the energies of ferromagnetic and antiferromagnetic ground states for these magnetic nanostructures. In addition, the study of the joint effect (***impurity+vacancy)-induced magnetization*** should be important for the understanding of the possible ways of controlled changes of magnetic properties of the nanomaterials.

**Acknowledgement**

Financial support of the RFBR (Grants 07-03-96061, 07-03-00026 and 08-08-00178) and Grant MK-502.2008.3 is gratefully acknowledged.

Table.
Magnetic moments (MMs, in $\mu_B$) for B, C and N doped and beryllium vacancy-containing beryllium monoxide nanotubes.

| * Wall defect | MMs (impurity) | **MMs $(Be)_{max}$ | MMs $(O)_{max}$ | MM (total, per cell) |
|---|---|---|---|---|
| *armchair* (6,6) BeO-NTs | | | | |
| B | 0.747 | 0.124 | 0.008 | 1.0 |
| C | 1.603 | 0.104 | 0.065 | 2.0 |
| N | 0.895 | 0.005 | 0.015 | 1.0 |
| vac Be | - | 0.016 | 0.630 | 2.0 |
| *zigzag* (10,0) BeO-NT | | | | |
| B | 0.732 | 0.175 | 0.021 | 1.0 |
| C | 1.585 | 0.118 | 0.026 | 2.0 |
| N | 0.897 | 0.006 | 0.016 | 1.0 |
| vac Be | - | 0.015 | 0.724 | 2.0 |

* both nanotubes with O vacancy are non-magnetic
** the maximal induced magnetic moments on the Be and O atoms in the nearest surrounding of the wall defects.

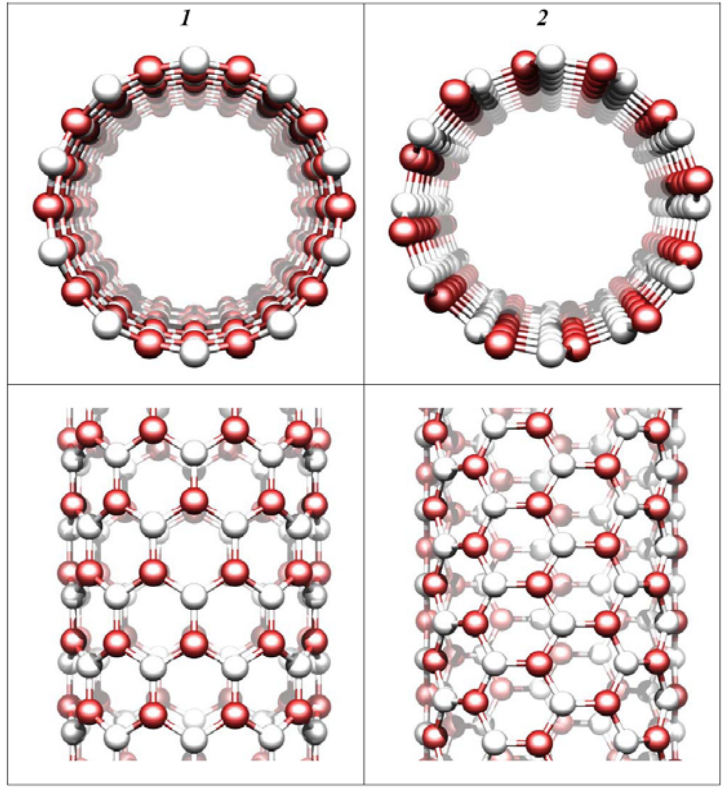

Fig. 1. (*Color online*) Top and side views of atomic structures of single-walled: 1.- *zigzag* (10,0) and 2- *armchair* (6,6) nanotubes of graphite-like hexagonal BeO.

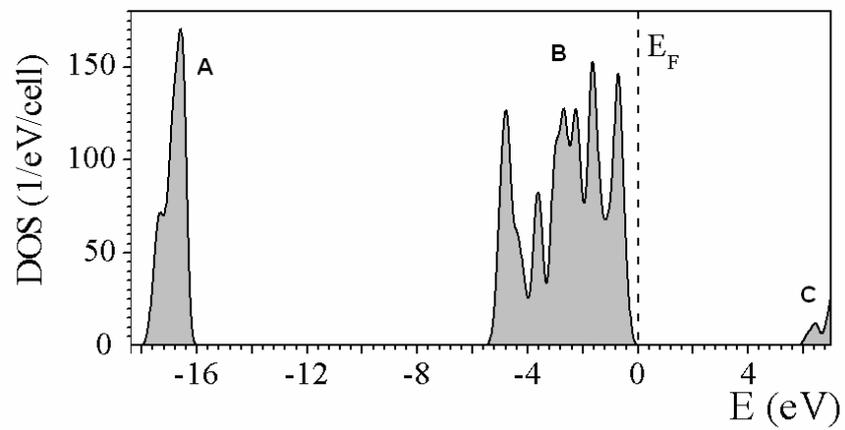

Fig. 2. Total density of states of pristine *armchair* (6,6) BeO nanotube.

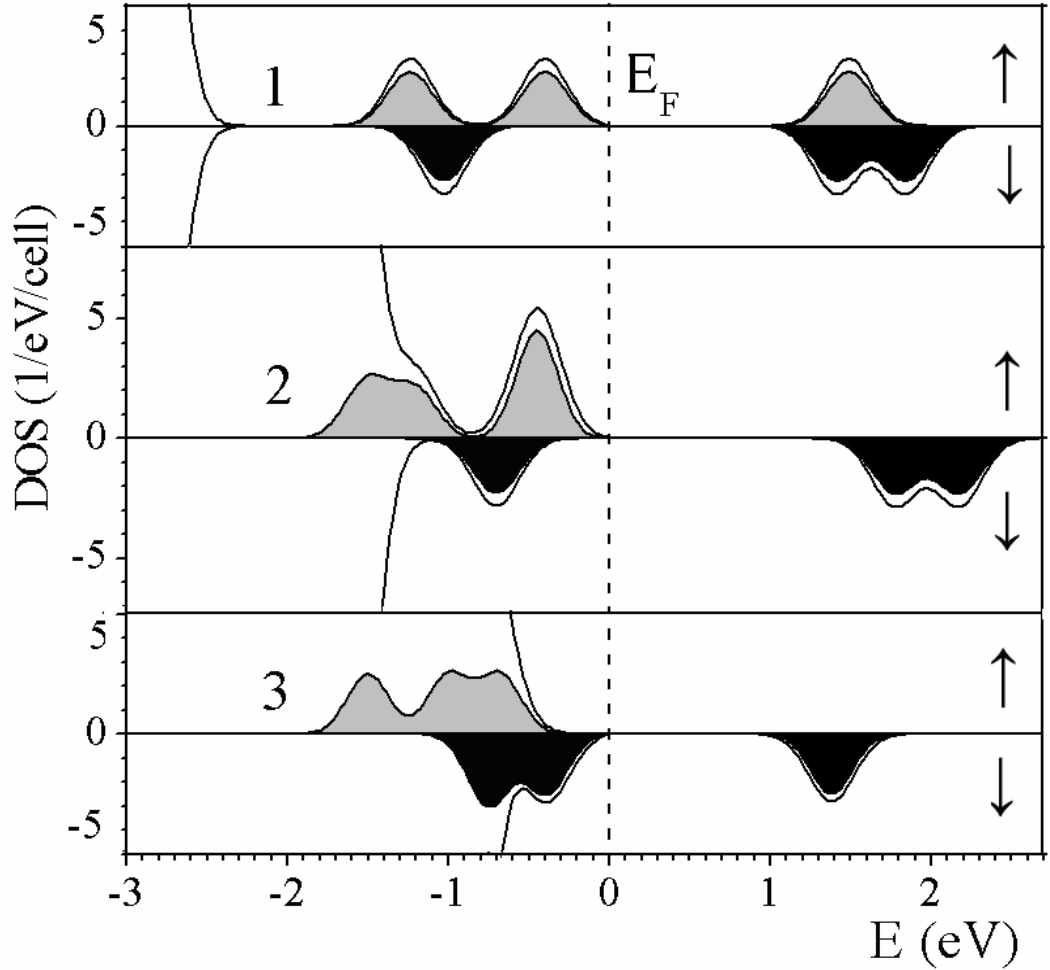

Fig. 3. Majority and minority total spin densities of states of a boron (1), carbon (2) and nitrogen-doped (3) *armchair* (6,6) BeO nanotube (*full lines*) and B, C and N impurity atoms $2p^{\uparrow\downarrow}$ states (*shaded*). The Fermi level is given at 0.0 eV.